\documentclass[preprint,12pt,authoryear]{elsarticle}
\usepackage{xcolor,amsmath,amssymb,amsfonts,array,graphicx}
\usepackage{hyperref,lmodern}
\hypersetup{colorlinks=false,allbordercolors=blue,pdfborderstyle={/S/U/W 1}}
\usepackage[utf8]{inputenc}
\usepackage[T1]{fontenc}
\usepackage[polish,english]{babel}
\usepackage{graphicx}
\usepackage{geometry,tabularx,array,booktabs}
\usepackage{enumitem}
\usepackage{placeins}
\usepackage{hyperref}
\AtBeginDocument{\selectlanguage{english}}

\journal{Fractals}

\begin{document}
	
	\begin{frontmatter}
		
		\title{The Binary Crisis Clock: \\Controlled by Sparse Ternary Interventions}
		
		\author{Ma\l gorzata Nowak-K\c epczyk} 
		
		\affiliation{organization={Institute of Informatics,\\ The John Paul II Catholic University of Lublin\\},
			addressline={Konstantynow 1H}, 
			city={Lublin},
			postcode={20-708}, 
			country={Poland, email: malnow@kul.pl}}

	\maketitle
	
\begin{abstract}
	We investigate time-dependent modular discrete-Laplacian automata on a
	triangular lattice, comparing purely binary evolution across different
	neighbourhood masks with dynamics subjected to a finite number of early
	ternary interventions.
	
	The corrected simulations reveal a common binary seed-return mechanism
	for all tested mask--seed combinations. Between successive return events,
	the configuration expands into a mask-dependent large-scale envelope,
	whereas at the characteristic times
	\[
	t=8k,\qquad k=1,2,3,\ldots,
	\]
	it collapses into spatially separated translated replicas of the initial
	seed. The mask determines the large-scale outline, while compatible
	seed--mask symmetries constrain the internal organization of the evolving
	configuration.
	
We then replace selected binary updates during the developmental
interval \(t\leq32\) by ternary updates and examine the subsequent
purely binary evolution. All schedules were enumerated for one to three
interventions, while \(5000\) unique schedules were sampled uniformly
for each \(n_3=4,\ldots,10\). Two to three well-placed interventions
produce most of the attainable density gain, increasing the mean density
over \(33\leq t\leq80\) from the binary reference value of approximately
\(0.068\) to about \(0.47\). Additional interventions increase the
highest observed density only marginally, but substantially enlarge the
fraction of schedules entering the high-density regime. Thus, timing is
critical in the very sparse regime, whereas additional interventions make
trajectory selection progressively more robust.
	
Ternary shaping does not abolish the binary crisis clock. All \(100\)
highest-ranked trajectories exhibited a detected crisis at
\(\tau=64\), measured from the final ternary intervention; in \(95\)
cases it exceeded the predefined major-crisis threshold and was the
first major post-intervention collapse. Intermediate crises were
substantially shallower than in the binary reference and random-control
trajectories.
\end{abstract}

\vspace{6mm}

\begin{graphicalabstract}\centering
\includegraphics[width=0.95\textwidth]{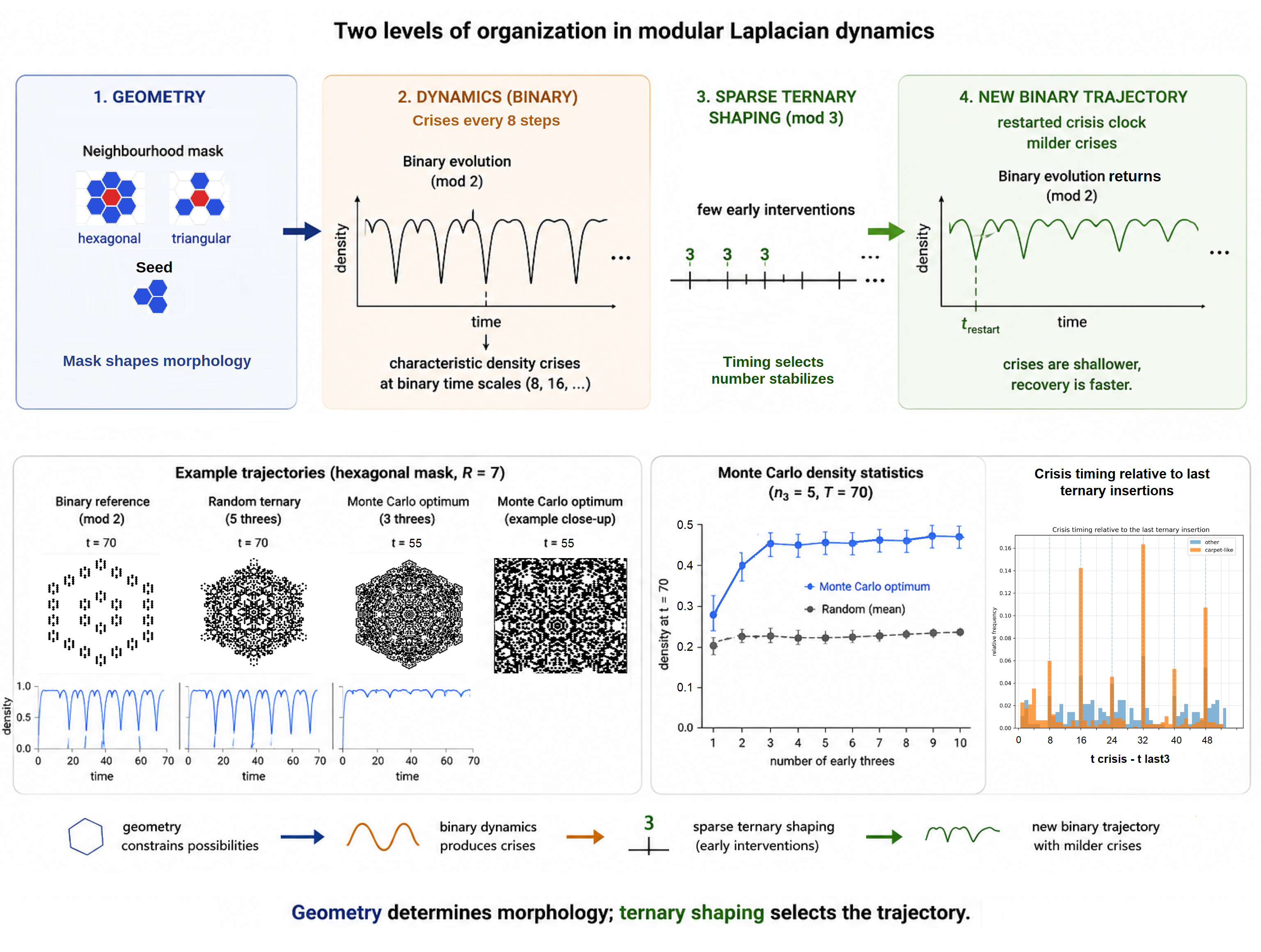}
\end{graphicalabstract}


\end{frontmatter}

\section{Introduction}

Additive and linear cellular automata provide a classical setting in
which simple local update rules generate geometrically complex but
partially algebraically tractable dynamics. Over finite fields and finite
rings, modular superposition and binomial-coefficient structure may produce
replication, recurrent motifs, and Sierpiński-type self-similar patterns
\cite{Martin1984,Takahashi1992,Prunescu2009}. Related work has connected
the fractal structure of cellular-automaton evolution with the algebraic
properties of the underlying state space and update operator
\cite{Gutschow2010}.

In two spatial dimensions, the geometry of the lattice and neighbourhood
can substantially affect growth directions, symmetry, and the large-scale
shape of evolving configurations \cite{PackardWolfram1985}. Fractal and
recurrent structures generated by two-dimensional additive rules have
also been studied in connection with Ulam-type automata and linear
elementary cellular automata \cite{Kawaharada2014}. More specifically,
iteration dynamical systems generated by discrete Laplacians on plane
lattices were investigated by Hadlich et al., who established fixed-point
and periodicity results and examined binomial and trinomial construction
sequences \cite{Hadlich2011}.

Previous studies of modular discrete-Laplacian dynamics on square lattices
revealed Fro\-be\-nius-type replication, recurrent density crises, and
large-scale seed-return events
\cite{NowakKepczyk2025Frobenius,MNK2026}. In the purely binary case, an expanding
configuration periodically decomposes into spatially separated translated
copies of its initial seed. Mixed binary--ternary schedules can redirect
this evolution and produce long-lived high-density transients
\cite{NowakKepczyk2025Stabilization}. These observations motivate the first problem considered
here: whether the binary seed-return mechanism survives a change from the
square lattice to a triangular lattice, and which aspects of the resulting
morphology are determined by the neighbourhood mask and initial seed.

The second problem concerns temporal intervention. In nonlinear dynamical
systems, finite perturbations may steer a trajectory toward a different
region of state space without permanently changing the governing dynamics
\cite{Cornelius2013}. We adopt an analogous strategy in a non-autonomous
modular cellular automaton. A small number of early binary updates is
replaced by ternary updates, after which the system again evolves
exclusively under the binary rule. The ternary steps are therefore treated
as finite developmental interventions rather than as persistent forcing.

The present study addresses four questions:

\begin{enumerate}
	\item Does the characteristic binary seed-return rhythm persist across
	different neighbourhood masks and seeds on the triangular lattice?
	
	\item How do mask geometry and seed--mask symmetry compatibility affect
	the large-scale envelope and internal organization of the evolving
	configurations?
	
	\item Can a small number of ternary interventions select a denser
	subsequent binary trajectory, and which intervention times are most
	effective?
	
	\item After the ternary phase has ended, how are the timing, depth, and
	recovery of the remaining binary density crises modified?
\end{enumerate}

Our computational results show that all tested masks retain the same
binary seed-return rhythm at \(t=8k\), while producing different
mask-dependent envelopes. Sparse ternary interventions do not remove this
binary mechanism. Instead, appropriately timed interventions select a
long-lived high-density binary trajectory, attenuate its intermediate
crises, accelerate recovery, and reset the timing of the first pronounced
post-intervention collapse.

\section{Methods}

\subsection{Modular Laplacian dynamics}

We consider a cellular automaton on a regular hexagonal tiling, equivalently
represented by the triangular lattice of cell centres. Let \(V\) denote the
set of lattice sites and let \(M\) be a finite neighbourhood mask expressed
as a set of lattice offsets.

For a seed \(S\subset V\), the initial state is
\[
\nu_0(p)=
\begin{cases}
	1, & p\in S,\\
	0, & p\notin S.
\end{cases}
\]

The modular Laplacian evolution is defined for \(t\geq1\) by
\[
\nu_t(p)=
\left[
\sum_{d\in M}
\bigl(\nu_{t-1}(p+d)-\nu_{t-1}(p)\bigr)
\right]\bmod k_t,
\]
where \(k_t\in\{2,3\}\). The summation is performed over the integers
before modular reduction.

\begin{figure}[h!] 
	\centering
	\includegraphics[width=0.7\textwidth]{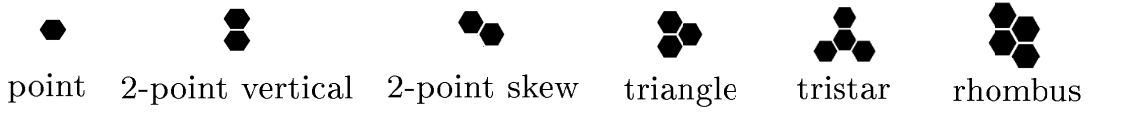}
	\caption{Seed configurations used in the experiments.}
	\label{fig:seeds}
\end{figure}

\begin{figure}[h!] 
	\centering
	\includegraphics[width=0.52\textwidth]{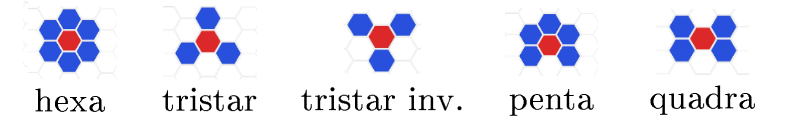}
	\caption{Neighbourhood masks used in the experiments.}
	\label{fig:neighborhoods}
\end{figure}

Purely binary evolution corresponds to \(k_t=2\) for every \(t\).
A ternary intervention at time \(t\) means that \(k_t=3\) in the update
\(\nu_{t-1}\mapsto\nu_t\); all non-intervention updates use \(k_t=2\).
The seed configurations and neighbourhood masks used in the experiments
are shown in Figures~\ref{fig:seeds} and~\ref{fig:neighborhoods}.

\subsubsection{Density}

Let
\[
A_t=\{p\in V:\nu_t(p)\neq0\}
\]
denote the set of active cells after \(t\) iterations. Both non-zero
ternary states, \(1\) and \(2\), are treated as active.

Let \(R_t\) be the set of lattice sites reachable from the initial seed
in at most \(t\) steps using the neighbourhood mask. Density is defined as
\[
\rho(t)=\frac{|A_t|}{|R_t|}.
\]

The denominator therefore depends only on the seed, neighbourhood mask,
and iteration number, and is independent of the intervention schedule.
For the one-point seed and the full hexagonal mask \texttt{hex6},
\[
|R_t|=1+3t(t+1).
\]

For an evaluation interval \(I\), we calculate
\[
\overline{\rho}(I)
=
\frac{1}{|I|}
\sum_{t\in I}\rho(t)
\]
and
\[
\sigma_\rho(I)
=
\sqrt{
	\frac{1}{|I|}
	\sum_{t\in I}
	\left(\rho(t)-\overline{\rho}(I)\right)^2
}.
\]

The term \emph{carpet-like} is used descriptively for an expanding,
high-density configuration with comparatively small fluctuations over
the specified observation window; it does not imply asymptotic or
permanent stability.

\subsection{Crisis detection and recovery}

A density crisis is identified as a sufficiently pronounced local minimum
of \(\rho(t)\). For a window of \(w\) preceding iterations, define
\[
\rho_{\mathrm{pre}}(t)
=
\frac{1}{w}\sum_{i=1}^{w}\rho(t-i),
\]
and let
\[
\sigma_{\mathrm{loc}}(t)
=
\operatorname{std}
\bigl(\rho(t-w),\ldots,\rho(t+w)\bigr).
\]

The normalized crisis score is
\[
K(t)
=
\frac{\rho_{\mathrm{pre}}(t)-\rho(t)}
{\sigma_{\mathrm{loc}}(t)+\varepsilon}.
\]

An iteration \(t\) is classified as a crisis when
\[
\rho(t)
=
\min_{s\in[t-w,t+w]}\rho(s)
\qquad\text{and}\qquad
K(t)>\kappa.
\]

For every detected crisis, its relative depth is
\[
D(t)
=
\frac{\rho_{\mathrm{pre}}(t)-\rho(t)}
{\rho_{\mathrm{pre}}(t)},
\]
and its recovery time is
\[
R(t)
=
\min\left\{
s>t:
\rho(s)\geq0.9\,\rho_{\mathrm{pre}}(t)
\right\}-t.
\]

The same parameters were used in all comparisons:
\[
w=4,\qquad
\varepsilon=10^{-9},\qquad
\kappa=1.0.
\]
A local minimum was retained only when its relative depth exceeded
\(0.05\). For a flat minimum, only its first iteration was recorded.
Only crises occurring after the final ternary intervention were included
in the post-intervention comparison. 

If the trajectory did not recover to \(90\%\) of the pre-crisis density
within the observation window, the recovery time was treated as censored.

\subsection{Search over ternary intervention schedules}

The schedule search was performed for the one-point seed and the full
hexagonal neighbourhood mask \texttt{hex6}. A ternary intervention
reported at displayed time \(t\) means that modulus \(3\) was used in the
transition
\[
\nu_{t-1}\longmapsto\nu_t.
\]
All other transitions were performed modulo \(2\).

Ternary interventions were restricted to the developmental interval
\[
T\subseteq\{1,\ldots,32\}.
\]
After \(t=32\), all trajectories evolved exclusively under the binary rule.

For schedules containing \(n_3=1,2,3\) ternary interventions, all
admissible combinations were enumerated:
\[
\binom{32}{1}=32,\qquad
\binom{32}{2}=496,\qquad
\binom{32}{3}=4960.
\]

For each \(n_3=4,\ldots,10\), \(5000\) unique schedules were sampled
uniformly without replacement from the corresponding set of admissible
combinations. No intervention times were preferentially weighted.
The pseudorandom sampling was performed using a fixed master seed, and
the exact seeds and computational settings are provided in the deposited
metadata.

Candidate schedules were evaluated over the post-developmental interval
\[
I_{\mathrm{eval}}=\{33,\ldots,80\}.
\]
For each trajectory, we calculated the mean density
\(\overline{\rho}_{33:80}\), its standard deviation
\(\sigma_{\rho,33:80}\), and the ranking score
\[
J=
\overline{\rho}_{33:80}
-
0.50\,\sigma_{\rho,33:80}.
\]

Higher values of \(J\) therefore identify trajectories combining high
mean density with comparatively small density fluctuations. The same
score was used to identify both the highest- and lowest-ranked schedules.

For \(n_3=1,2,3\), the highest-ranked schedule is the optimum within the
complete search space. For \(n_3\geq4\), the reported schedule is the
highest-ranked schedule observed in the uniform sample and is not claimed
to be the global optimum.

\subsection{Post-intervention crisis comparison}

For the post-intervention crisis analysis, the highest-ranked and
random-control groups were matched with respect to the number of ternary
interventions \(n_3\). For each \(n_3\) stratum, the control group
contained the same number of trajectories as the highest-ranked group
and was sampled uniformly from the remaining schedules. This yielded
\(100\) highest-ranked trajectories and \(100\) matched random controls.

The selected trajectories were extended to \(t=160\). Only crises
occurring after the final ternary intervention were included. A major
crisis was defined by relative depth \(D\geq0.40\).

\section{Results}

To address the questions posed above, we perform three experiments. We first classify binary dynamical regimes on triangular lattices, then investigate sparse ternary interventions as a mechanism for trajectory shaping, and finally analyse the timing and severity of density crises after the intervention phase.

\subsection{Experiment 1. Binary dynamics for different masks and seeds}

The corrected simulations revealed a common binary mechanism across all
tested seed--mask combinations. Purely binary evolution alternates between
growth and abrupt seed-return events. During the growth phase, the
configuration develops a large-scale envelope whose geometry is determined
by the neighbourhood mask. At the characteristic binary times
\[
t=8k,\qquad k=1,2,3,\ldots,
\]
the expanded structure collapses and the seed-return stage is recovered.

\begin{table}[p] 
	\centering
	\small
	\setlength{\tabcolsep}{4pt}
	\renewcommand{\arraystretch}{1.15}
	
	\begin{tabularx}{\linewidth}{|
			>{\centering\arraybackslash}p{4cm}|
			>{\centering\arraybackslash}X|
			>{\centering\arraybackslash}X|
			>{\centering\arraybackslash}p{4cm}|}
		\hline
		
		\textbf{Neighbourhood mask}
		&	\textbf{Pre-crisis state, $t=31$}
		&	\textbf{Replication event, $t=32$}
		&	\textbf{Inherited large-scale outline}	\\
		\hline
		
		\begin{tabular}[c]{@{}c@{}}
			\includegraphics[width=0.8cm]{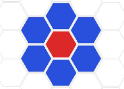}\\
			hex6
		\end{tabular}
		&
		\includegraphics[width=0.95\linewidth,height=1.5cm,keepaspectratio
		]{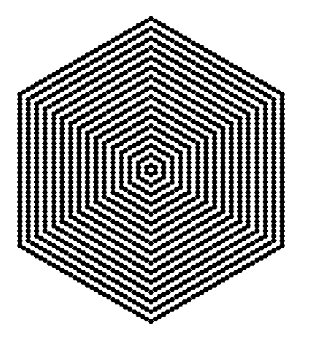}
		&
		\includegraphics[width=0.95\linewidth,height=1.5cm,	keepaspectratio
		]{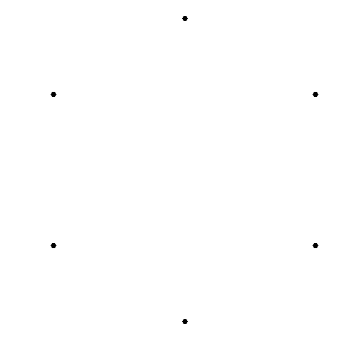}
		&
		Hexagonal
		\\
		\hline
		
		\begin{tabular}[c]{@{}c@{}}
			\includegraphics[width=0.78cm]{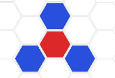}\\
			tristar
		\end{tabular}
		&
		\includegraphics[width=0.95\linewidth,
		height=1.5cm,keepaspectratio
		]{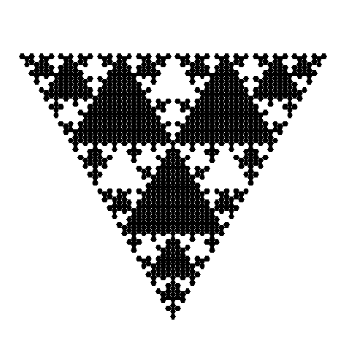}
		&
		\includegraphics[width=0.95\linewidth,
		height=1.5cm,keepaspectratio
		]{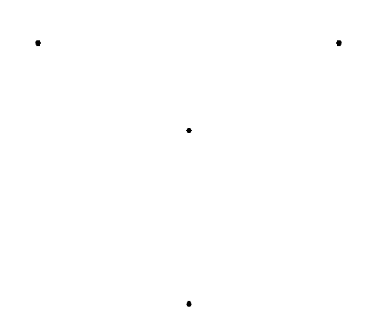}
		&
		Triangular, Sierpiński-like
		\\
		\hline
		
		\begin{tabular}[c]{@{}c@{}}
			\includegraphics[width=0.62cm]{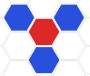}\\
			tristar inverted
		\end{tabular}
		&
		\includegraphics[width=0.95\linewidth,
		height=1.5cm,keepaspectratio
		]{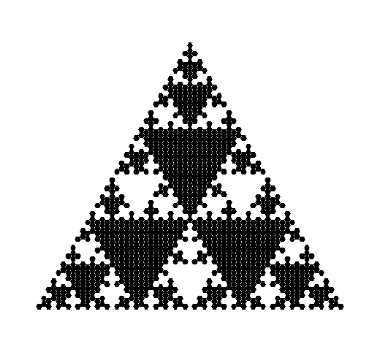}
		&
		\includegraphics[width=0.95\linewidth,
		height=1.5cm,keepaspectratio
		]{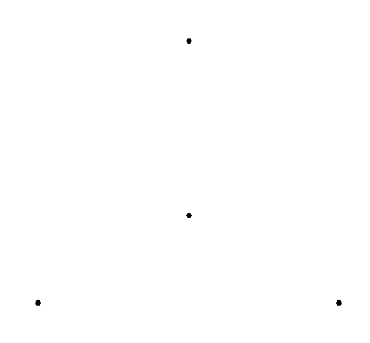}
		&
		Mirrored triangular, Sierpiński-like
		\\
		\hline
		
		\begin{tabular}[c]{@{}c@{}}
			\includegraphics[width=0.75cm]{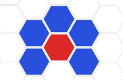}\\
			penta
		\end{tabular}
		&
		\includegraphics[width=0.95\linewidth,
		height=1.5cm,keepaspectratio
		]{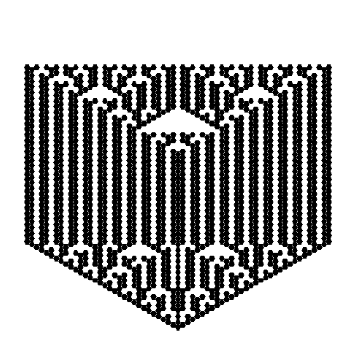}
		&
		\includegraphics[width=0.95\linewidth,
		height=1.5cm,keepaspectratio
		]{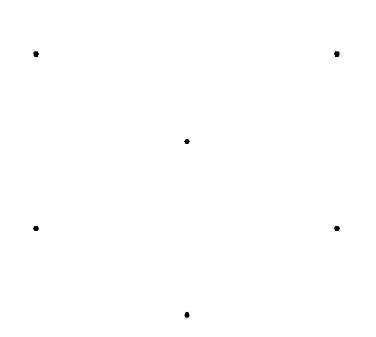}
		&
		Asymmetric pentagonal or skew
		\\
		\hline
		
		\begin{tabular}[c]{@{}c@{}}
			\includegraphics[width=0.72cm]{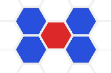}\\
			quadra
		\end{tabular}
		&
		\includegraphics[width=0.95\linewidth,
		height=1.5cm,keepaspectratio
		]{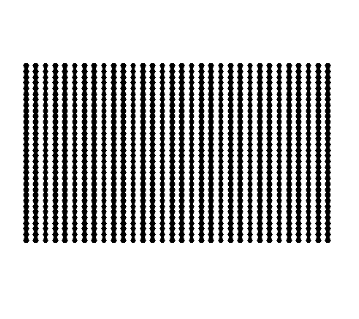}
		&
		\includegraphics[width=0.95\linewidth,
		height=1.5cm,keepaspectratio
		]{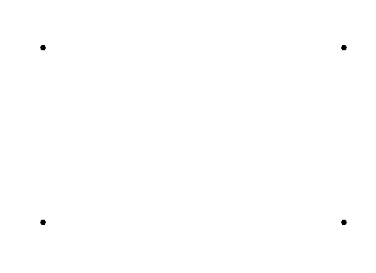}
		&
		Quadrilateral or rectangular
		\\
		\hline
		
	\end{tabularx}
	
	\caption{%
		Mask-dependent geometry of purely binary modular Laplacian
		dynamics for the one-point seed. In every tested case, the
		configuration first expands into a large-scale structure whose
		outline is inherited from the neighbourhood mask. Immediately
		before the characteristic binary return time, illustrated here at
		$t=31$, the support forms a fully developed mask-shaped figure.
		At $t=32$, the structure collapses and the seed-return stage is
		recovered. The same return mechanism recurs at the characteristic
		binary times $t=8k$, $k=1,2,3,\ldots$. The triangular masks
		additionally produce pronounced Sierpiński-like triangular
		organization during the pre-return growth phase.
	}
	\label{tab:binary_mask_outlines}
\end{table}

The masks therefore determine the geometry of the expanding envelope rather
than the qualitative dynamical regime. The full hexagonal mask produces
hexagonal outlines, the quadra mask produces quadrilateral structures, and
the asymmetric penta mask generates skew envelopes. Both triangular masks
produce triangular structures with a pronounced Sierpiński-like internal
organization. Representative pre-return and return states for the one-point
seed are shown in Table~\ref{tab:binary_mask_outlines}.


\subsubsection{Compatibility of seed and mask symmetries}

Although the neighbourhood mask determines the large-scale envelope, the
symmetry of the internal pattern depends on the compatibility between the
mask and the initial seed. Let
\[
G_M=\{g:\ gM=M\}
\]
denote the symmetry group of the neighbourhood mask and let
\[
G_S=\{g:\ gu_0=u_0\}
\]
denote the symmetry group of the initial configuration, both regarded as
groups of transformations acting on the same embedded triangular lattice.

For every transformation \(g\in G_M\), the corresponding Laplacian operator
commutes with \(g\):
\[
L_M(gu)=gL_M(u).
\]
Consequently, every symmetry belonging simultaneously to the mask and the
seed is preserved throughout the evolution:
\[
G_M\cap G_S\subseteq G_{u_t},
\qquad t\geq 0.
\]

The abstract type of the symmetry groups is not sufficient. Their axes and
centres must coincide in the actual lattice embedding. When a reflection
axis of the seed coincides with a reflection axis of the mask, the generated
figures retain that axial symmetry. When the reflection axes do not coincide,
the common symmetry subgroup may be reduced to a half-turn, a threefold
rotation, a single reflection, or even the identity.

This explains the variety observed among geometrically equivalent seeds.
For the centrally symmetric \texttt{quadra} mask, differently oriented
two-cell and rhombic seeds generate distinct zigzag motifs, but a common
half-turn symmetry may remain. For the \texttt{tristar\_rev} mask and the
skew rhombus, one compatible reflection axis survives. In contrast, the
combination of the \texttt{tristar} mask with the triangular seed preserves
a threefold rotational symmetry, even though their reflection axes do not
coincide.

Additional symmetries may occasionally appear at particular iterations as
a result of modular cancellations. Such symmetries are accidental
time-dependent properties of a configuration and are not guaranteed by the
common symmetry subgroup.

\begin{figure}[h] 
	\centering
	\includegraphics[width=0.8\textwidth]{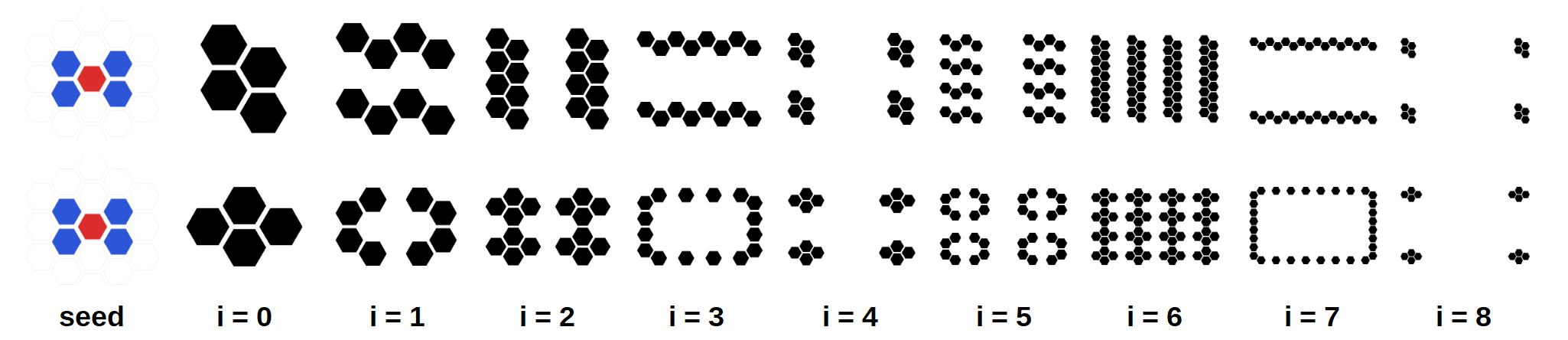}
	
	\caption{%
		Influence of seed orientation and symmetry-axis compatibility on binary
		evolution for the asymmetric \texttt{penta} mask. The seeds
		\texttt{rhombus} and \texttt{rhombus\_skew} are equivalent up to a
		lattice rotation, but their reflection axes occupy different positions
		relative to the mask. Consequently, the two orientations generate
		different internal structures. The large-scale envelope remains
		constrained by the mask, whereas the symmetry of the evolving
		configuration is determined by the spatially aligned transformations
		common to the seed and the mask.
		\label{fig:seed_mask_symmetry}
	}
\end{figure}

The effect of spatial alignment is illustrated in
Figure~\ref{fig:seed_mask_symmetry}, where two rotationally equivalent
rhombic seeds generate different internal structures under the same
asymmetric mask.

\paragraph{Binary crisis rhythm on square and triangular lattices} 

The symmetry effects described above accompany a more fundamental binary
mechanism. The recurrent seed-return dynamics is identical to that
previously observed on the square lattice \cite{MNK2026}. In both lattice
geometries, the configuration expands between successive return times and,
at
\[
t=8k,\qquad k=1,2,3,\ldots,
\]
undergoes an abrupt density collapse into spatially separated translated
copies of the initial seed.

We refer to this regular sequence of density collapses and seed-return
events as the \emph{binary crisis rhythm}. The associated phase structure,
which determines the timing of successive crises, will be termed the
\emph{binary crisis clock}. Changing the underlying lattice geometry
modifies the mask-dependent large-scale outlines, but does not alter the
binary seed-return cycle. Within each cycle, the mask determines the
large-scale envelope, the seed determines the replicated motif, and the
common spatial symmetry subgroup \(G_M\cap G_S\) determines which
symmetries are necessarily preserved.

\subsection{Experiment 2: Sparse ternary shaping}

Experiment~1 showed that all tested neighbourhood masks exhibit the same
purely binary seed-return mechanism, with recurrent collapses at
\[
t=8k,\qquad k=1,2,3,\ldots.
\]
We selected the full hexagonal mask \texttt{hex6} as the principal test
system because its sixfold symmetry produces a particularly clear and
reproducible reference trajectory, while minimizing additional effects
associated with seed orientation and mask asymmetry. The \texttt{quadra}
mask is considered separately as a transparent example of recurrent motif
redirection.

Motivated by the effects of mixed modular dynamics previously observed on
the square lattice \cite{MNK2026}, we investigate whether a
small number of ternary updates can redirect the subsequent binary
trajectory. Ternary steps are treated as finite developmental
interventions: after the last intervention, all subsequent updates are
again performed modulo~2.

\subsubsection{Does timing matter?}

\begin{figure}[h] 
	\centering
	\includegraphics[width=0.9\textwidth]{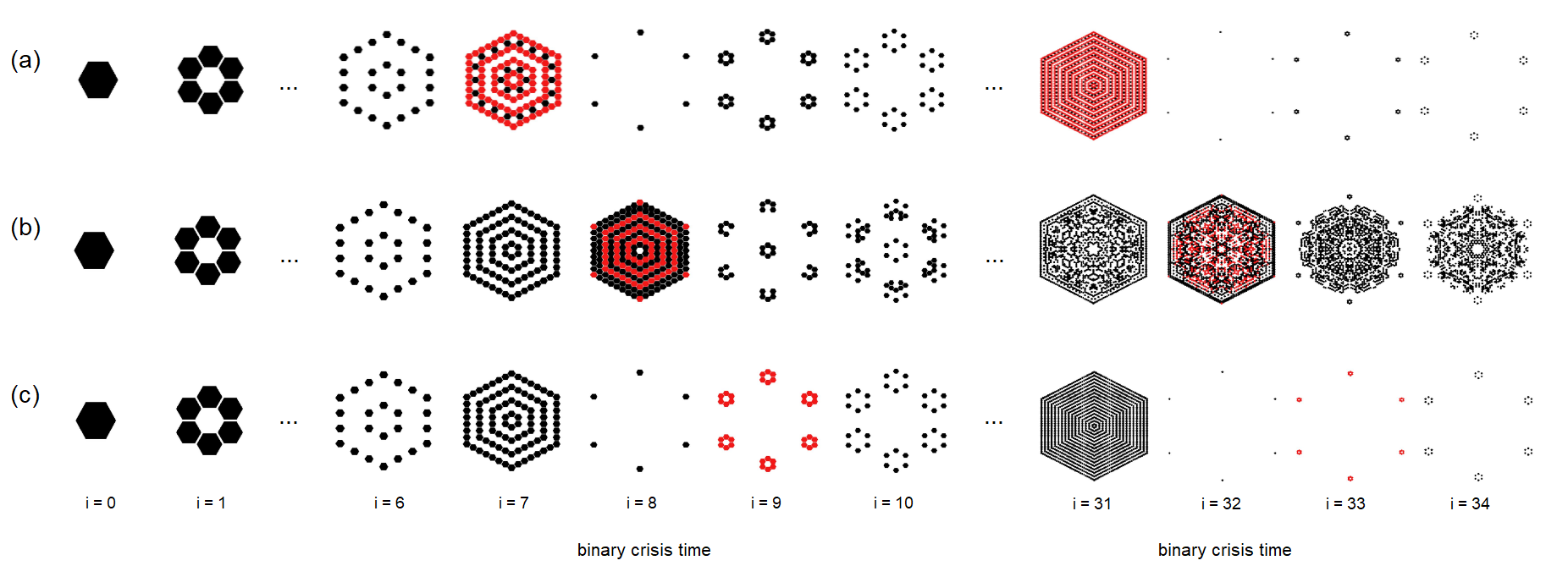}
\caption{%
	Influence of intervention timing relative to the binary seed-return
	cycle for the one-point seed and the full hexagonal mask
	\texttt{hex6}. Ternary updates were applied periodically at
	(a) every \(t=8k-1\), (b) every \(t=8k\), and (c) every \(t=8k+1\),
	respectively, while all remaining updates were binary. Representative
	states are shown near two consecutive seed-return events. Red cells
	indicate states immediately produced by ternary updates. Periodic ternary updates applied at every \(t=8k\) redirected the
	subsequent evolution toward a markedly denser trajectory, whereas schedules
	shifted by one iteration produced evolutions much closer to the binary
	reference cycle.
}
\label{8k_interventions}
\end{figure}

We compared ternary insertions applied at
$t=8k-1$, $8k$, and $8k+1$. The resulting evolutions proved highly
sensitive to timing. Interventions synchronized with the characteristic
binary crisis rhythm ($t=8k$) generated substantially denser
carpet-like structures, whereas shifts by a single iteration produced
trajectories resembling the binary reference evolution
(Figure~\ref{8k_interventions}).

\subsubsection{What does a ternary insertion change?}

The timing experiment demonstrates that ternary interventions can strongly
influence the subsequent binary evolution. We now ask a more fundamental
question: which aspect of the trajectory is modified by a single ternary
update?

\begin{figure}[h!] 
	\centering
	\includegraphics[width=0.85\textwidth]{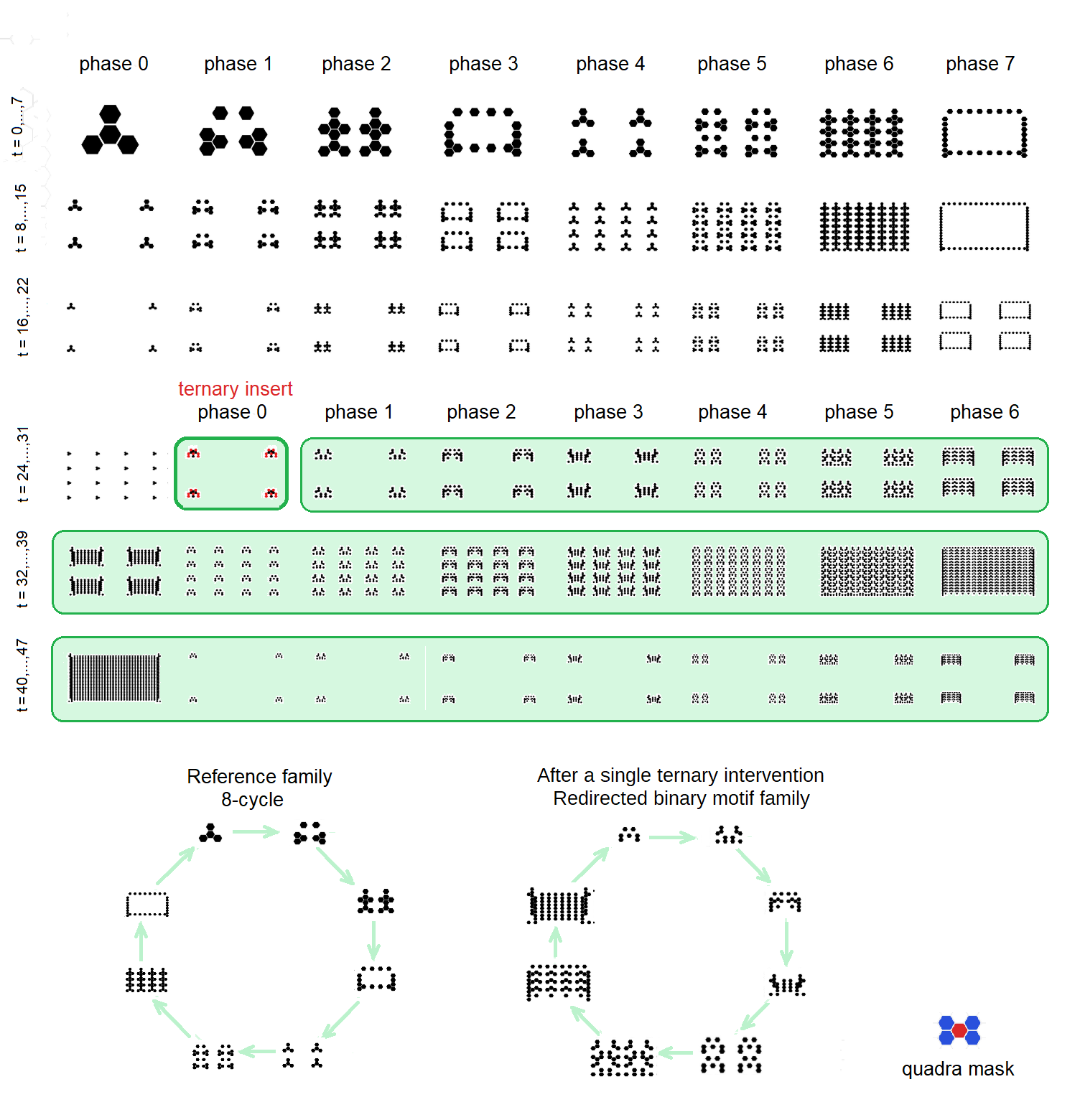}
	
	\caption{%
		Redirection of a recurrent binary motif hierarchy by a single
		ternary intervention for the \texttt{quadra} mask. The purely
		binary reference trajectory follows an eight-phase hierarchy, with
		related motifs recurring in successive intervals between the
		characteristic crisis times \(t=8k\). At \(t=25\), one binary
		update is replaced by a ternary update; all subsequent updates are
		again performed modulo~2. Nevertheless, the later trajectory does
		not recover the original reference family but follows a visibly
		different recurrent sequence of motifs. The lower diagrams compare
		the reference eight-phase family with the family observed after
		the ternary intervention.
		\label{fig:self_sim}
	}
\end{figure}

For the \texttt{quadra} mask, purely binary evolution follows a recurrent
eight-phase hierarchy of self-similar motifs. Figure~\ref{fig:self_sim}
shows the effect of replacing the binary update at \(t=25\) with a single
ternary update. All subsequent updates are again performed modulo~2.
Nevertheless, the trajectory does not return to the original reference
family. Instead, it follows a visibly different sequence of recurrent
motifs.

Thus, the ternary intervention does not merely modify one configuration.
It changes the state from which the subsequent binary evolution proceeds
and thereby redirects the system toward a different recurrent motif
family. In this sense, a single ternary update acts as a developmental
switch between distinct binary trajectories.

\subsubsection{Can sparse ternary interventions select denser binary
	trajectories?}

The preceding experiments show that a ternary update may act as a
developmental switch rather than as a transient local perturbation.
We therefore investigated how the number and timing of early ternary
updates affect the subsequent purely binary trajectory.

All intervention times were restricted to \(t\leq32\). Schedules with
one to three ternary interventions were exhaustively enumerated, while
for each \(n_3=4,\ldots,10\), \(5000\) unique schedules were sampled
uniformly. Every schedule was ranked using its mean density and density
fluctuation over \(33\leq t\leq80\), as described in the Methods.

Intervention times were sampled without replacement from the set
\(\{1,\ldots,32\}\).


\subsubsection{Experiment 2 results}

The schedule search revealed two complementary effects: a rapid increase
in the highest attainable density in the very sparse regime, and a
progressive increase in the robustness of high-density trajectory
selection as the number of interventions increased.

\paragraph{A small number of interventions produces the principal density gain}

\begin{figure}[t] 
	\centering
	\includegraphics[width=\textwidth]
	{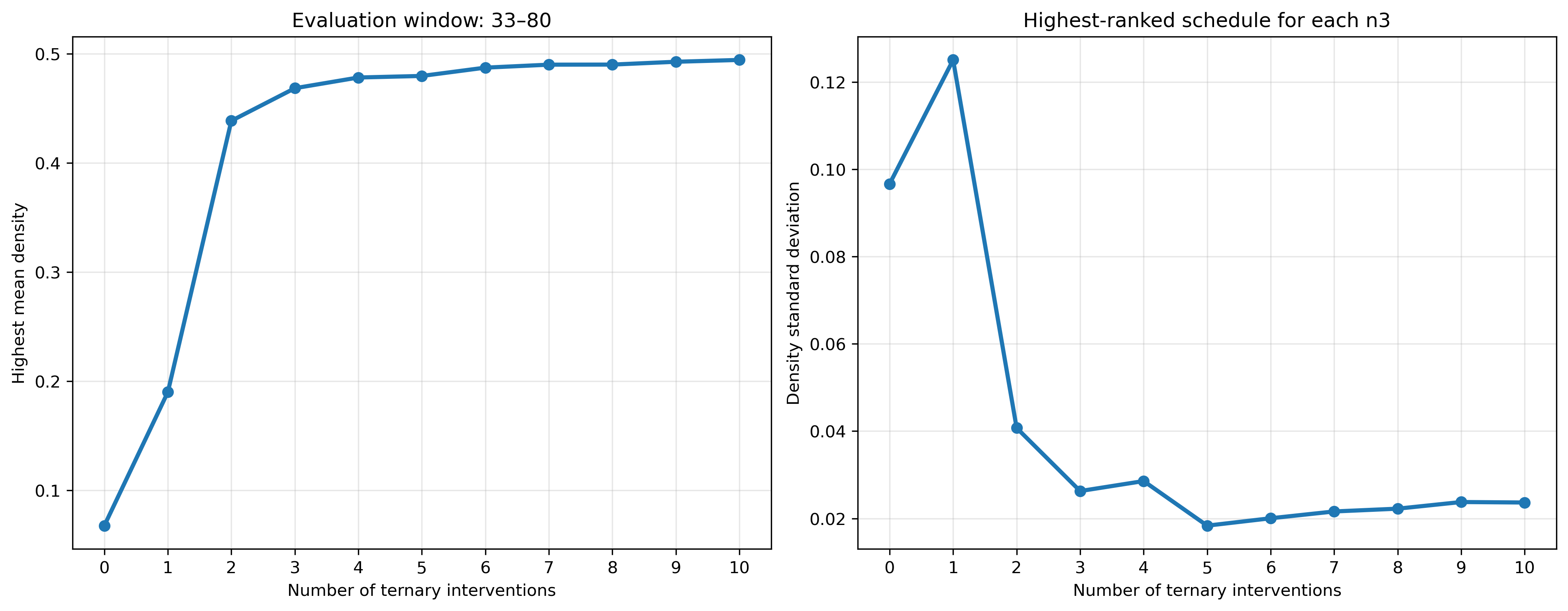}
	\caption{
		Highest-ranked density outcomes as a function of the number of
		ternary interventions. The left panel shows the highest observed
		mean density over \(33\leq t\leq80\); the right panel shows the
		corresponding density standard deviation for the same schedule.
		The binary reference is shown at \(n_3=0\).
		For \(n_3=1,2,3\), all admissible schedules were enumerated.
		For each \(n_3=4,\ldots,10\), the point represents the
		highest-ranked schedule among \(5000\) uniformly sampled unique
		schedules. The rapid increase for two to three interventions is
		followed by a broad density plateau.
	}
	\label{fig:best_vs_n3}
\end{figure}

The largest change occurred between one and two ternary interventions.
The highest mean density increased from approximately \(0.190\) for
\(n_3=1\) to \(0.438\) for \(n_3=2\), and reached approximately \(0.468\)
for \(n_3=3\). For \(n_3\geq3\), the highest observed densities formed a
broad plateau near \(0.47\)--\(0.49\)
(Figure~\ref{fig:best_vs_n3}).

Thus, two or three well-placed ternary updates are sufficient to produce
most of the maximum density gain observed over the investigated
evaluation window. Additional interventions increase the highest
observed density only marginally.

The reduction in density fluctuations is similarly pronounced. The
highest-ranked trajectories with three or more ternary interventions
typically exhibit density standard deviations of approximately
\(0.02\)--\(0.03\), compared with substantially larger fluctuations in
the purely binary and single-intervention trajectories.

\begin{table}[h] 
	\centering
	\small
	\begin{tabular}{crrcrr}
	\hline
	$n_3$ & Lowest-ranked schedule & $\bar\rho$ & Highest-ranked schedule & $\bar\rho$ & $N$ \\
	\hline
	0 & -- & 0.068 & -- & 0.068 & 1 \\
	1 & 31 & 0.068 & 16 & 0.190 & 32 \\
	2 & 29,31 & 0.068 & 8,23 & 0.438 & 496 \\
	3 & 27,29,31 & 0.068 & 16,17,30 & 0.468 & 4960 \\
	4 & 3,7,13,31 & 0.068 & 16,17,21,32 & 0.478 & 5000 \\
	5 & 7,11,19,23,31 & 0.068 & 8,16,17,19,30 & 0.480 & 5000 \\
	6 & 2,11,13,17,18,29 & 0.068 & 2,3,17,24,25,29 & 0.487 & 5000 \\
	7 & 2,7,10,15,17,23,25 & 0.068 & 10,11,17,22,25,28,32 & 0.490 & 5000 \\
	8 & 9,11,13,19,21,23,29,31 & 0.068 & 12,13,16,18,24,26,27,29 & 0.490 & 5000 \\
	9 & 1,13,17,19,23,25,26,29,31 & 0.068 & 9,10,11,14,16,17,28,29,32 & 0.493 & 5000 \\
	10 & 2,6,9,11,13,17,19,21,23,25 & 0.068 & 8,9,10,13,20,21,25,26,28,29 & 0.494 & 5000 \\
	\hline
\end{tabular}
	\caption{
		Highest- and lowest-ranked intervention schedules under
		\(J=\overline{\rho}-0.50\sigma_\rho\), evaluated over
		\(33\leq t\leq80\).
		For \(n_3=1,2,3\), all admissible schedules were enumerated;
		for \(n_3=4,\ldots,10\), \(5000\) unique schedules were sampled
		uniformly. Schedules were ranked using \(J\); mean density is shown
		for interpretability.
	}
	\label{tab:master_best_worst}
\end{table}

\paragraph{Timing remains critical in the sparse regime}

\begin{figure}[h!] 
	\centering
	\includegraphics[width=\textwidth]
	{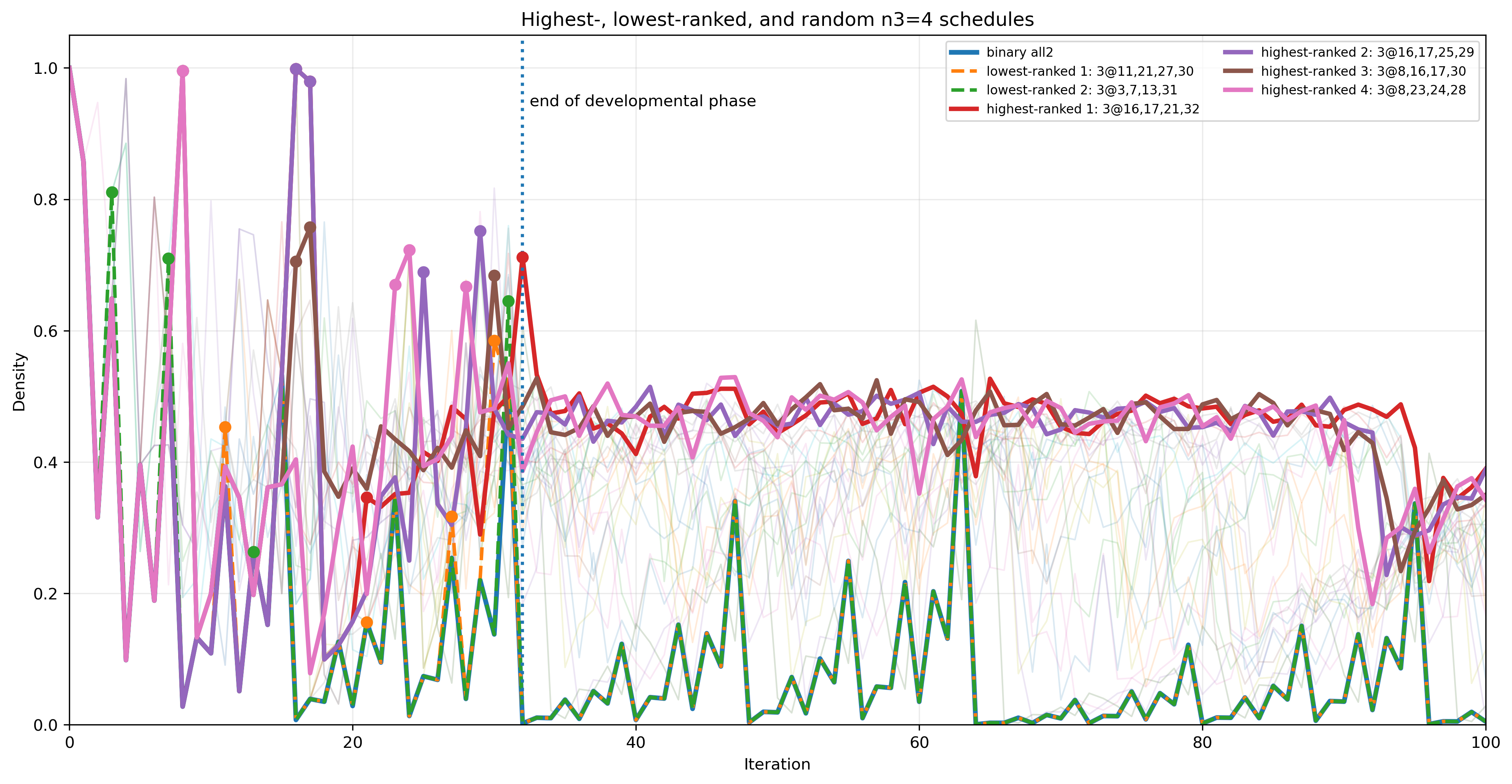}
	\caption{
		Density trajectories for selected four-intervention schedules.
		Thick solid curves show the four highest-ranked schedules, dashed
		curves show the two lowest-ranked schedules, and pale curves show
		\(30\) uniformly selected control schedules. The purely binary
		reference is shown for comparison. Filled markers indicate the
		displayed times of ternary interventions, and the vertical dotted
		line marks the end of the developmental phase at \(t=32\).
		Despite containing the same number of ternary updates, the
		lowest-ranked schedules remain close to the binary crisis trajectory,
		whereas the highest-ranked schedules enter a long-lived high-density
		regime.
	}
	\label{fig:selected_trajectories}
\end{figure}

Schedules containing the same number of ternary interventions may produce
markedly different outcomes. For four interventions, the lowest-ranked
trajectories remain close to the oscillatory binary reference evolution,
whereas the highest-ranked schedules enter a long-lived high-density
regime with \(\rho(t)\) typically near \(0.45\)--\(0.50\) over much of
the evaluation interval
(Figure~\ref{fig:selected_trajectories}).

This contrast shows that the number of interventions alone does not
determine the result. In the sparse regime, their placement remains
essential: poorly timed schedules may be dynamically almost ineffective,
even when they contain the same number of ternary updates as successful
ones.

\begin{figure}[t] 
	\centering
	\includegraphics[width=0.9\textwidth]
	{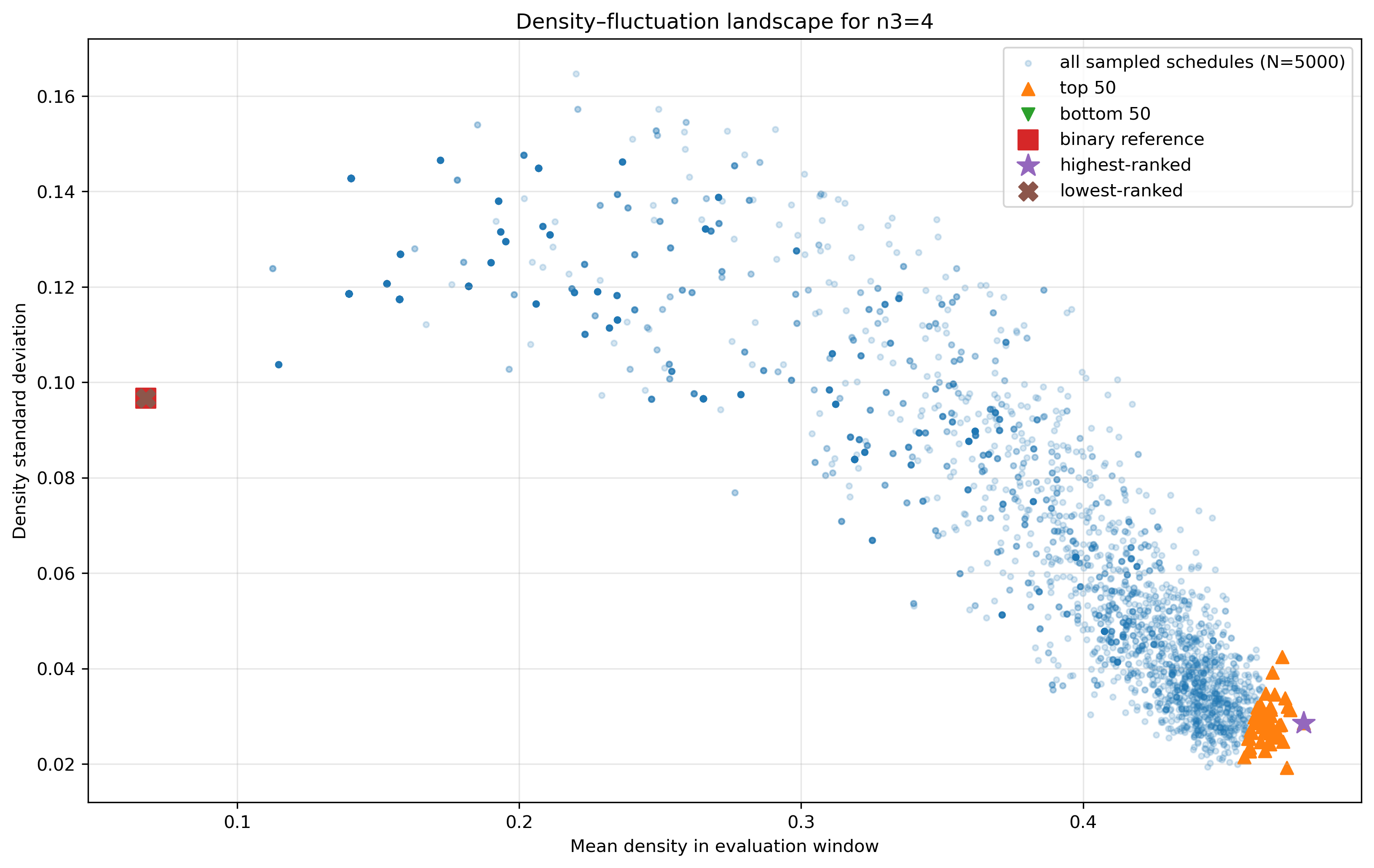}
	\caption{
		Mean post-intervention density versus density fluctuation for
		\(5000\) uniformly sampled schedules containing four ternary
		interventions. Each point represents one schedule evaluated over
		\(33\leq t\leq80\). The highest- and lowest-ranked schedules were
		identified using \(J=\overline{\rho}-0.50\sigma_\rho\).
	}
	\label{fig:n3_4_scatter}
\end{figure}

The density--fluctuation landscape for \(n_3=4\) confirms that these
outcomes form a continuous but strongly structured distribution
(Figure~\ref{fig:n3_4_scatter}). Most high-density schedules occupy a
compact low-fluctuation region, whereas binary-like schedules are
characterized by both lower mean density and larger temporal variability.

\paragraph{Additional interventions increase robustness rather than the
	highest observed density}

\begin{figure}[h!] 
	\centering
	\includegraphics[width=0.9\textwidth]
	{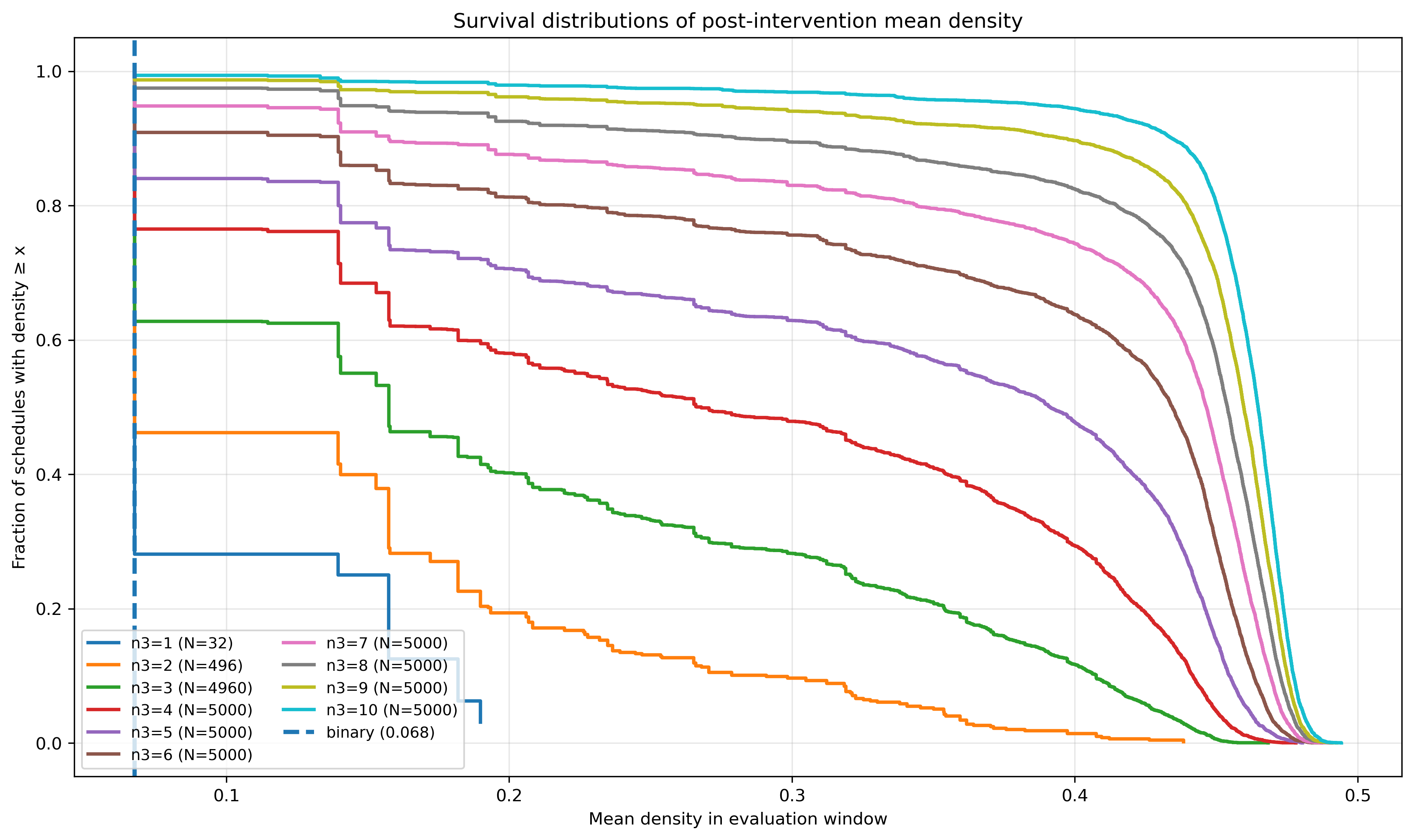}
	\caption{
		Survival distributions of mean post-intervention density for
		increasing numbers of ternary interventions. For a threshold \(x\),
		each curve gives the fraction of schedules satisfying
		\(\overline{\rho}_{33:80}\geq x\).
		The spaces for \(n_3=1,2,3\) were exhaustively enumerated;
		the curves for \(n_3=4,\ldots,10\) are based on \(5000\) uniformly
		sampled unique schedules for each \(n_3\).
		The progressive rightward shift shows that additional interventions
		primarily increase the robustness and probability of high-density
		trajectory selection rather than substantially increasing the
		highest observed density.
	}
	\label{fig:density_survival}
\end{figure}

Although the highest observed density changes little beyond three to four
interventions, the distribution of outcomes changes strongly.
The survival distributions in Figure~\ref{fig:density_survival} shift
progressively toward higher density as \(n_3\) increases. Consequently,
a large post-intervention density becomes increasingly common among
uniformly sampled schedules.

The principal effect of additional ternary interventions is therefore
not a continued increase in the highest observed density, but an
enlargement of the set of schedules leading to the high-density regime.
Precise timing is critical when ternary interventions are very sparse;
with additional interventions, successful trajectory selection becomes
progressively more robust to the exact schedule.

Taken together, the results indicate a transition from timing-sensitive
sparse control to a broader high-density basin of intervention schedules.
A few well-placed updates are sufficient to reach the dense regime,
whereas additional updates mainly increase the probability of reaching it.

The intervention phase therefore does not permanently replace the binary
rule. It selects a new state from which the subsequent evolution proceeds
entirely modulo \(2\). The next experiment examines whether the resulting
high-density trajectories suppress the binary crisis mechanism or instead
modify its phase, depth, and recovery dynamics.


\subsection{Experiment 3. Post-intervention binary dynamics:
	crisis-clock resetting and crisis attenuation}

Experiment~2 showed that a small number of developmental ternary
interventions can redirect crisis-prone binary evolution toward
long-lived high-density carpet-like trajectories. We now examine the
dynamics after the intervention phase has ended and all subsequent
updates are again performed modulo~2.

The central question is whether ternary shaping suppresses the binary
crisis mechanism or instead changes its phase and severity.

\subsubsection{Crisis timing}

\begin{figure}[ht] 
	\centering
	\includegraphics[width=0.9\textwidth]
	{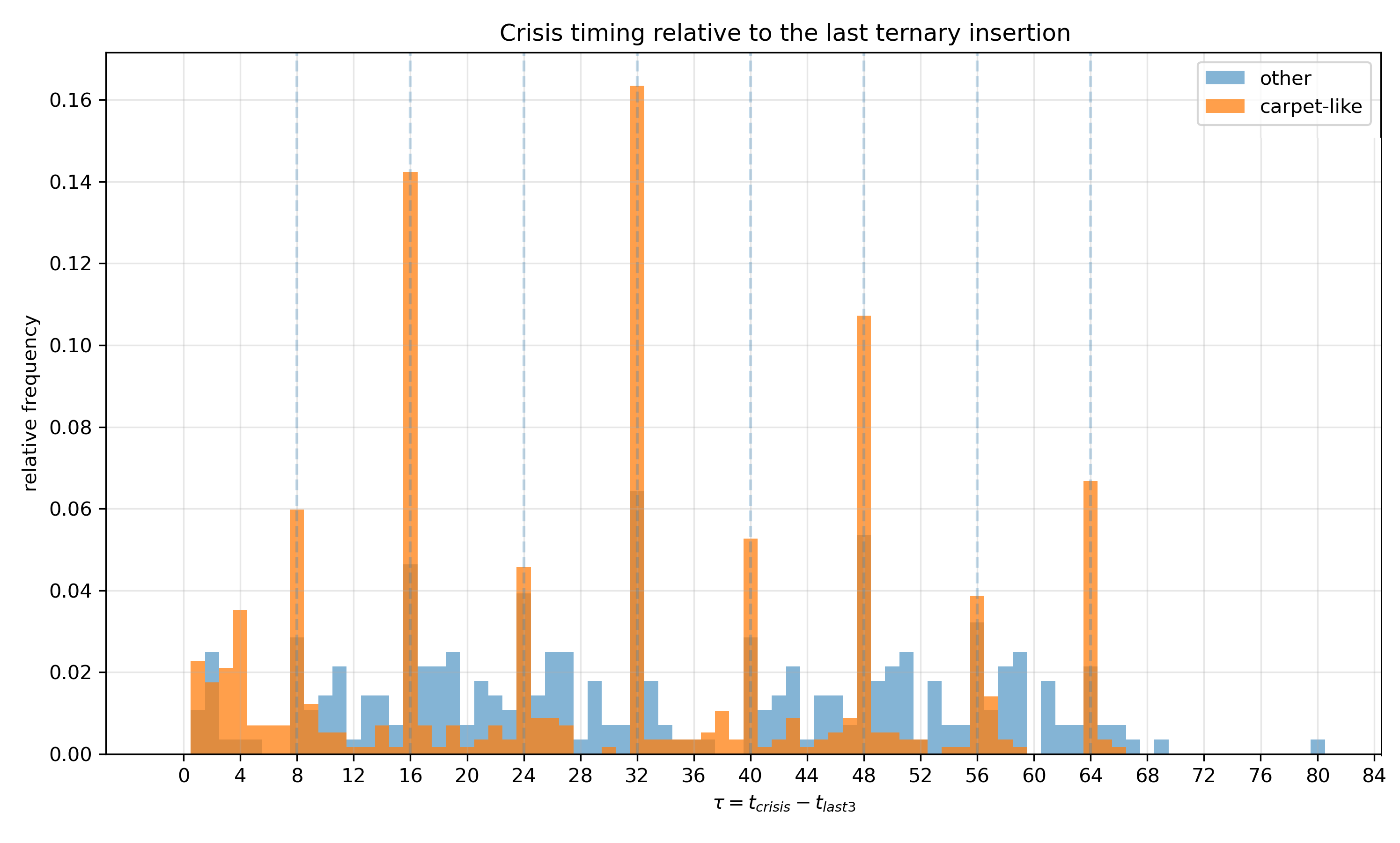}
\caption{
	Distribution of post-intervention crisis times measured relative to
	the final ternary intervention,
	\(\tau=t_{\mathrm{crisis}}-t_{\mathrm{last3}}\), for
	\(100\) highest-ranked trajectories and \(100\) random controls.
	Vertical dashed lines mark multiples of eight binary iterations.
	The highest-ranked trajectories exhibit a strong concentration of
	events at the characteristic binary phases, including the prominent
	event at \(\tau=64\).
}
	\label{fig:tau_histogram}
\end{figure}

To characterize the temporal organization of the post-intervention
dynamics, we measured each crisis time relative to the final ternary
intervention:
\[
\tau=t_{\mathrm{crisis}}-t_{\mathrm{last3}}.
\]
The distribution of detected crisis times relative to the final ternary
intervention is shown in Figure~\ref{fig:tau_histogram}.
The highest-ranked trajectories exhibit pronounced concentrations near
multiples of eight, with particularly strong peaks at
\(\tau=16\), \(32\), \(48\), and \(64\).

The crisis rhythm remains visible after the developmental phase.
Shallow intermediate crises recur near
\[
\tau=8,16,24,\ldots,
\]
but they are followed by rapid recovery and do not immediately destroy
the high-density structure. The first pronounced collapse of the
high-density trajectories occurs at
\[
\tau=64,
\]
that is, exactly 64 binary iterations after the final ternary
intervention.

Thus, ternary shaping does not eliminate the binary crisis rhythm.
Instead, the timing of subsequent crises becomes organized relative to
the final ternary update. The binary crisis clock is therefore reset:
its eight-step rhythm survives, while the first major post-intervention
collapse is postponed until \(\tau=64\).

\subsubsection{Crisis severity}

The strongest distinction between the trajectory groups concerned crisis
severity. The median relative crisis depth was \(0.095\) in the
highest-ranked trajectories, compared with \(0.161\) in the
\(n_3\)-matched random controls and \(0.965\) in the binary reference.
The corresponding 90th percentiles were \(0.430\), \(0.599\), and
\(0.996\), respectively.

The mean number of detected post-intervention crises was also lower in
the highest-ranked trajectories (\(9.13\)) than in the matched random
controls (\(10.40\)); the binary reference exhibited \(15.00\) detected
crises. Median recovery time was one iteration in both intervention
groups, compared with five iterations in the binary reference.

Thus, after matching for the number of ternary interventions, the
highest-ranked schedules were still characterized by shallower and
slightly less frequent crisis events than random controls. Their median
recovery time, however, was unchanged.

Table~\ref{tab:master_crisis_summary} summarizes the  differences
between the binary reference trajectory, randomly sampled ternary
schedules, and the highest-ranked schedules.

\begin{table}[t] 
	\centering
	\small
	\begin{tabular}{lrrrrr}
		\hline
		Group & $N$ & Mean crises/traj. & Median depth & P90 depth & Median recovery  \\
		\hline
binary & 1 & 15.00 & 0.965 & 0.996 & 5.0  \\
random control & 100 & 10.40 & 0.161 & 0.599 & 1.0 \\
highest ranked & 100 & 9.13 & 0.095 & 0.430 & 1.0  \\
		\hline
	\end{tabular}
	\caption{
		Post-intervention crisis statistics for the binary reference,
		randomly sampled controls, and highest-ranked intervention schedules.
		The binary row represents one reference trajectory; the other rows
		summarize \(100\) trajectories each. A major crisis was defined by
		relative depth \(D\geq0.40\).
	}
	\label{tab:master_crisis_summary}
\end{table}

The crisis rhythm remains visible after the developmental phase, with
events concentrated near multiples of eight
(Figure~\ref{fig:tau_histogram}).

All \(100\) highest-ranked trajectories exhibited a detected crisis at
\(\tau=64\). In \(95\) cases, its relative depth exceeded the predefined
major-crisis threshold \(D\geq0.40\), and it was the first major
post-intervention crisis. In the remaining five trajectories, the crisis
at \(\tau=64\) remained below this threshold.

Thus, ternary shaping does not eliminate the binary crisis clock.
Instead, the post-intervention crisis sequence becomes organized relative
to the final ternary update, with a particularly consistent event at
\(\tau=64\).


\section{Conclusions}

The present work shows that modular Laplacian dynamics on triangular
lattices is governed by two complementary mechanisms: lattice geometry
determines the spatial form of the evolution, whereas sparse developmental
perturbations select among the binary trajectories compatible with that
geometry.

At the geometric level, all tested neighbourhood masks and seed
configurations exhibited the same fundamental binary crisis mechanism.
The expanding configuration repeatedly undergoes density collapse and
decomposes into translated replicas of the initial seed at the
characteristic times
\[
t=8k.
\]
The neighbourhood mask primarily determines the large-scale outline and
preferred directions of growth, while the seed influences the initial
symmetry and transient arrangement of the replicated components. Thus,
changing the mask modifies the geometry of the resulting patterns but does
not remove the underlying binary seed-return rhythm.

This result extends the binary crisis mechanism previously observed on
square lattices to triangular-lattice dynamics. The lattice geometry
changes the visible morphology of the configurations, whereas the
replication process and its characteristic temporal organization remain
closely related.

At the dynamical level, sparse ternary updates act as developmental
interventions. Two to three well-placed updates produce most of the
observed increase in post-intervention density. Beyond this sparse
regime, additional interventions increase the highest observed density
only slightly, but strongly enlarge the fraction of schedules entering
a high-density trajectory. Timing is therefore critical when
interventions are few, whereas increasing their number makes successful
trajectory selection progressively more robust.

Ternary interventions do not permanently stabilize the system. After
the developmental phase, all subsequent updates are binary and the
eight-step crisis organization remains detectable. The highest-ranked
trajectories exhibit markedly shallower post-intervention crises than
random controls and the binary reference. A crisis at \(\tau=64\) was
detected in every highest-ranked trajectory and constituted the first
major post-intervention collapse in \(95\%\) of them. Ternary shaping
therefore resets and attenuates the binary crisis clock rather than
abolishing it.

The most striking dynamical result is therefore that ternary shaping does
not eliminate the binary crisis clock. Instead, it resets its phase and
changes the severity of its subsequent events. Crisis times become
organized relative to the final ternary intervention, while the
intermediate collapses are attenuated and followed by more rapid recovery.

Sparse ternary perturbations consequently do not replace the binary rule.
They produce a new effective initial state from which the subsequent
purely binary dynamics proceeds. Their principal role is to select a more
favourable binary trajectory: one characterized by higher medium-term
density, shallower intermediate crises, faster reconstruction, and a
reset timing of the first major collapse.

More broadly, the results suggest a separation between spatial and
developmental control in modular Laplacian evolution. Geometry determines
the family of admissible morphologies, whereas sparse, temporally localized
interventions select among the corresponding dynamical trajectories. This
interpretation is consistent with broader approaches to pattern formation
and self-organization in which structural constraints and developmental
history play complementary roles
\cite{Haken2006,Turing1952,Whitesides2002}.

\subsection*{Open questions}

Several questions remain open. First, a complete analytical explanation
of the eight-step binary crisis rhythm on triangular lattices is still
missing. In particular, it remains to be determined how the observed
seed-return times follow from the algebraic structure of the modular
Laplacian operator and its Frobenius-type replication mechanism.

Second, the origin of the post-intervention time scale
\[
\tau=64
\]
requires further investigation. It is not yet known why the first
pronounced collapse of the ternary-shaped high-density trajectories occurs
exactly 64 binary iterations after the final ternary update, or whether
this delay can be predicted directly from the intervention schedule and
the resulting effective binary seed.

Further work should also examine the dynamics beyond the first major
post-intervention collapse. The present analysis establishes that the
ternary-shaped trajectories are long-lived but not permanent; it does not
classify the subsequent reconstruction, possible secondary resets, or
later generations of the crisis hierarchy.

A further problem concerns the relationship between mask geometry, seed
symmetry, and the organization of successful intervention windows.
Spectral, multiscale, and algebraic analyses may help determine why
particular crisis-adjacent schedules, such as interventions near
\(8k\) and \(8k+1\), redirect the binary dynamics more effectively than
other schedules.

Finally, it remains unknown whether sparse developmental control is a
specific feature of modular Laplacian evolution or a more general
principle of discrete self-organizing systems. Extending the analysis to
other modular rules, neighbourhoods, lattice geometries, and independently
validated time horizons may clarify the scope of the mechanism.

\section*{Acknowledgements}

The authors would like to acknowledge the inspiration provided by a long-standing line of investigations on binary and ternary structures developed by O. Suzuki, the late J. Ławrynowicz, and collaborators \cite{Lawrynowicz2018,Suzuki2022}. Over many years, these studies repeatedly highlighted the special role of interactions between the moduli 2 and 3 in a wide
variety of mathematical and physical contexts. The present work continues this line of investigation in the context of modular Laplacian dynamics.

\section*{Declarations}

\begin{itemize}
	\item Funding: Not applicable.
	\item Conflicts of Interest: The author declares no conflicts of interest.
	\item Author Contributions: Ma\l gorzata Nowak-K\c epczyk conducted all research and analysis.
	\item Data Availability: The code used to generate the experiments is available from the author upon request.
\end{itemize}

\appendix

\section{Preferred intervention windows}
\begin{table}[t] 
	\centering
	\small
	\begin{tabular}{lrrr}
		\hline
		Motif & Observed & Expected & Enrichment \\
		\hline
		(8,9) & 72 & 16.23 & 4.44 \\
		(16,17) & 77 & 16.23 & 4.74 \\
		(24,25) & 25 & 16.23 & 1.54 \\
		(8,16,17) & 16 & 3.32 & 4.82 \\
		(8,16,24,32) & 2 & 0.64 & 3.11 \\
		\hline
	\end{tabular}
\caption{
	Descriptive enrichment of selected ternary-intervention motifs
	among the \(50\) highest-ranked schedules for each
	\(n_3=4,\ldots,10\). Expected counts were calculated under uniform
	placement with the same intervention counts. The results are descriptive
	and no post-selection significance tests are reported.
}
	\label{tab:master_motif_enrichment}
\end{table}

\bibliographystyle{plainnat}  

\bibliography{references}

\end{document}